\documentclass[preprint,aps,prb,showpacs]{revtex4}

\usepackage{graphicx}% Include figure files
\usepackage{dcolumn}% Align table columns on decimal point
\usepackage{bm}% bold math

\begin{document}
\title{Charge and spin distributions in GaMnAs/GaAs Ferromagnetic Multilayers}
\author{S. C. P. Rodrigues, L. M. R. Scolfaro,  J. R. Leite}
\author{I. C. da Cunha Lima}
\altaffiliation[On leave from ]{Instituto de F\'\i sica,
Universidade do Estado do Rio de Janeiro, Rio de Janeiro, R.J.,
Brazil}
\affiliation{Instituto de F\'\i sica, Universidade de S\~ao Paulo,\\
CP 66318, 05315-970, S\~ao Paulo, SP, Brazil}
\author{G. M. Sipahi}
\affiliation{Instituto de F\'\i sica de S\~ao Carlos,
Universidade de
S\~ao Paulo, \\
CP 369, 13560-970, S\~ao Carlos, SP, Brazil}
\author{M. A. Boselli }
\affiliation{Departamento de F\'\i sica, Universidade Federal de Ouro Preto\\
Campus Universitário Morro do Cruzeiro, 35.400-000 Ouro Preto,
M.G., Brazil}

\date{\today}

\begin{abstract}
A self-consistent electronic structure calculation based on the
Luttinger-Kohn model is performed on GaMnAs/GaAs multilayers. The
Diluted Magnetic Semiconductor layers are assumed to be  metallic
and ferromagnetic. The high Mn concentration (considered as 5\% in
our calculation) makes it possible to assume the density of
magnetic moments as a continuous distribution, when treating the
magnetic interaction between holes and the localized moment on the
Mn$^{++}$ sites. Our calculation  shows the distribution of heavy
holes and light holes in the structure.  A strong
spin-polarization is observed, and the charge is concentrated
mostly on the GaMnAs layers, due to heavy and light holes with
their total angular momentum aligned anti-parallel to the average
magnetization. The charge and spin distributions are analyzed in
terms of their dependence on the number of multilayers, the widths
of the GaMnAs and GaAs layers, and the width of lateral GaAs
layers at the borders of the structure.

\end{abstract}

\pacs{73.21.-b, 75.75.+a, 85.75.-d,72.25.Dc}

\maketitle

\section{Introduction}
\label{Intro}

Recent advances on the physics and technology of GaAs-based
nanostructures with diluted magnetic semiconductors (DMS)  open a
wide range of potential applications of these systems  in
integrated magneto-optoelectronic devices. \cite{nat} In
Ga$_{1-x}$Mn$_{x}$As alloys substitutional Mn acts as an acceptor
(it binds one hole), and at the same time it carries a localized
magnetic moment, due to its five electrons in the 3 \textit{d}
shell. For $x$ near 0.05, the alloy is a metallic ferromagnet,
\cite{matsukura} the Curie-Weiss temperature after annealing is
160K, \cite{newohno} and the free hole concentration is near
$10^{20-21}cm^{-3}$. The possibility of application in Spintronics
and Photonics exists because such a layer provides the injection
of spin-polarized carriers in an otherwise non-magnetic
semiconductor region of the device, eliminating the needs of a
strong external magnetic field. Obviously, the higher the
transition temperature obtained for a DMS thin layer, the higher
the possibilities for a device with such a layer to operate near
room temperatures. The ferromagnetic order in the metallic phase
is understood, at present, as resulting from the indirect exchange
between the Mn$^{2+}$ ions due to the local spin polarization of
the hole gas. This explanation implies the spin coherence length
to be larger than the average distance of the localized moments.
Although most of the scientific theoretical effort in this problem
has been directed to understand the origin of the ferromagnetism,
there are still fundamental issues to be considered in the
electronic structure. For instance, the role played by the light
holes and holes of the split-off band is a point yet to be
understood. Such a nomenclature in GaAs is specific for bulk
systems, where the tetragonal symmetry is preserved. In the case
of heterostructures, where such symmetry is broken by the presence
of the interfaces, a mixing occurs in the composition of the hole
states. Differently from heavy holes, light hole states are not
spin-eigenstates. In consequence, the occurrence of a local
magnetic field is also a factor that contributes to the mixing in
the composition of the hole states. This can be seen easily in the
framework of the effective mass approximation and Luttinger-Kohn
(LK) \textbf{k}$\cdot$\textbf{p} expansion, \cite{kohn} because
off-diagonal terms appear in the Hamiltonian, breaking the
tetragonal symmetry. Therefore, the presence of interfaces,
together with a local magnetic field, claims for a better
treatment of the calculation of the electronic properties in DMS
heterostructures.

To the present, the six bands \textbf{k}$\cdot$\textbf{p} method
has been used to obtain the valence band structure of (Ga,Mn)As
only in bulk systems. \cite{dietl,lee,vurgaf,abolf} Some
calculations included the effects of biaxial strain, spin-orbit
coupling, and exchange correlation in a parabolic band
approximation. \cite{dietl1,kim,jungwir} In the case of quantum
wells, (Ga,Mn)As multilayers and superlattices, a self-consistent
calculation has been performed for parabolic heavy holes subbands.
\cite{luc} Self-consistent calculations have also been performed
in Refs. \onlinecite{kim} and \onlinecite{jungwir} assuming
isotropic effective masses. To the exception of  Ref.
\onlinecite{macdon}, where a Monte Carlo simulation is used, the
electronic structure calculation assumes a homogeneous density of
magnetic moment, as well as homogeneous negative charge
concentration (due to the ionized Mn atoms). Relaxing  this
approximation implies in considering a multiple-scattering
treatment, what is outside the scope of the present work. The
homogeneous approximation for the density of magnetic and Coulomb
scattering centers (localized magnetic moments and ionized
impurities) provides important information concerning the carriers
charge and spin distributions.

Here we present a self-consistent LK \textbf{k}$\cdot$\textbf{p}
calculation for GaMnAs/GaAs multilayers and superlattices. As
described below, we adopt a super-cell calculation which is an
extension of the LK method to treat the cases of quantum wells and
superlattices (SL). The structure we consider  consists of
substitutional Mn ions uniformly distributed in
Ga$_{0.95}$Mn$_{0.05}$As layers of width $d_1$, with a hole
concentration equivalent, in bulk, to $1$x$10^{20}$ cm$^{-3}$,
assumed to be metallic and ferromagnetic, at T=0K.  The DMS  are
separated by non-magnetic GaAs layers of width $d_2$. Before the
first DMS layer and after the last one, GaAs lateral layers of
width $s$ complete the structure, as shown in Fig. \ref{fig1}. The
super-cell model consists of placing this ``active part'',
described so far, between thick layers of a large gap material,
assumed here to be AlAs, and treating the whole system as a
superlattice.

The final spin and charge configuration contains a complete
information about the composition of heavy holes and light holes
in the structure. Carriers are anti-parallel heavy holes, and
anti-parallel light holes, in a lesser amount. A strong spin
polarization is observed, and the charge is concentrated mostly on
the GaMnAs layers. The charge and spin distributions are analyzed
in terms of their dependence on the number of multilayers, the
widths $d_1$ and $d_2$, as well as the widths of the lateral GaAs
layers.

\section{Magnetic Interaction in the confined LK  model}
\label{Vmag}

The interaction between free holes and the localized magnetic
moments  is well described by the Kondo-like term
\begin{equation}
V_{mag}({\bf{r}})= -I\sum_i \vec s({\bf{r}})\cdot
{\bf{S}}({\bf{R}}_i)\delta (\vec{r}-{\bf{R}}_i), \label{Kondo}
\end{equation}
where $I$ is the $sp-d$ interaction. The localized spin of the Mn
ion $\vec S_i$ at position $\vec R_i$ is treated as a classical
variable, since it results of the five 3$d$ electrons obeying
Hund's rule, and  no hybridization with carriers is considered,
due to the high difference in energies.  $\vec s(\vec r) $ is the
spin operator of the carrier at position $\vec r$. At zero
temperature, assuming a complete alignment of the localized
magnetic moment, i.e., ${\bf{S}}{(\bf{R}}_i)=\bf{S}$, we have:
\begin{equation}
V_{mag}(\vec{r})= -I {\bf{S}}\cdot\vec s(\vec r)\sum_i\delta
(\vec{r}-{\bf{R}}_i)=-\frac{I}{2}
{\bf{S}}\cdot\vec\sigma\rho_i(\vec{r}).
\end{equation}
In the last expression we used
$\vec\sigma=\hat{i}\sigma_x+\hat{j}\sigma_y+\hat{k}\sigma_xz$ to
denote the three Pauli matrices;  $\rho_i(\vec{r})$ is the
density of magnetic impurities. Assuming a homogeneous
distribution of the localized magnetic dipoles inside the DMS
layers , we have $\rho_i(\vec{r})\approx xN_0g(z)$, where $N_0$ is
the the density of cations, $x$ is the substitutional
concentration of Mn, and $g(z)=1$ if $z$ lies inside a DMS layer,
$g(z)=0$ otherwise. In that case the magnetic interaction becomes:
\begin{equation}
V_{mag}(z)=-\frac{x}{2}N_0\beta g(z)\vec{M}\cdot\vec{\sigma}.
\label{konsimp}
\end{equation}
In Eq.(\ref{konsimp})  we have explicitly taken into consideration
that carriers are holes, replacing $I$ by $\beta$. For GaMnAs,
$N_0\beta=-1.2$ eV. \cite{okaba} For electrons, $N_0\alpha=0.2$
eV.\cite{alfa} Were the spin of the particles well defined, this
term would represent, in bulk, a shift on the top (bottom) of the
valence (conduction) band. This is not the case, as discussed
above, for light holes and split-off holes in GaAs.

It is well known that the valence bands in GaAs at the $\Gamma$
point split, due to spin-orbit interaction, into four $j=3/2$
states belonging to the $\Gamma_8$ representation, and two $j=1/2$
states belonging to the $\Gamma_7$ representation. These two are
separated from the $\Gamma_8$ states by the spin-orbit energy
$\Delta$, which is 340 meV in GaAs. \cite{wu} We adopt the
notation $\mid j,m_j>$ to represent the $\Gamma_8$ and the
$\Gamma_7$ states, making use of the fact that these states are
also eigenstates of the total angular momentum operator
${\bf{J}}={\bf{L}}+\vec{s}$, with eigenvalue $j$, and
simultaneously eigenstates of its z-component, $J_z$,
corresponding to the eigenvalue $m_j$. Notice that the $\Gamma_8$
states $\mid v1>$ and $\mid v2>$ both having $j=3/2$ but with
$m_j=3/2$ and $m_j=-3/2$, respectively, are called heavy holes
states, having their spins well defined, being ``up'', i.e.,
aligned with ${\bf{J}}$, and ``down'', anti-aligned. They can be
represented by using as a basis the three \textit{p}-type states
in the directions $x,y,$ and $z$: \cite{Ender}
\begin{eqnarray}
\mid v1>&=& \, \,
\mid\frac{3}{2}\frac{3}{2}>=\frac{1}{\sqrt{2}}(\mid x +iy
\uparrow>)\\
\mid v2>&=&
\mid\frac{3}{2}\frac{\bar{3}}{2}>=\frac{i}{\sqrt{2}}(\mid x-iy
\downarrow>)
\end{eqnarray}
The other states, the light hole states $\mid v3>$ and $\mid v4>$
corresponding also to $j=3/2$, but with $m_j=1/2$ and $m_j=-1/2$,
and the split-off states $\mid v5>$ and $\mid v6>$ corresponding
to $j=1/2$ with $m_j=1/2$ and $m_j=-1/2$, do not have well defined
spins:
\begin{eqnarray}
\mid v3>&=& \mid\frac{3}{2}\frac{1}{2}>=\frac{i}{\sqrt{6}}(-2\mid
z \uparrow>+ \mid x+iy\downarrow>)\\
\mid
v4>&=&\mid\frac{3}{2}\frac{\bar{1}}{2}>=\frac{1}{\sqrt{6}}(\mid
x-iy\uparrow> + 2\mid z\downarrow>)\\
\mid v5>&=&\mid\frac{1}{2}\frac{1}{2}>=\frac{1}{\sqrt{3}}(\mid z
\uparrow> + \mid x+iy\downarrow>)\\
\mid
v6>&=&\mid\frac{1}{2}\frac{\bar{1}}{2}>=\frac{i}{\sqrt{3}}(-\mid
x-iy \uparrow> + \mid z\downarrow>)
\end{eqnarray}

The kinetic, Hartree, and the exchange-correlation terms appear in
this formalism as the well known components of the LK matrix.
\cite{kohn} In multilayers and superlattices the mismatches of the
valence and conduction bands, which play the roles of confining
potentials, are to be added. Differences on the lattice parameters
introduce the additional terms of the strain. Here we have also to
introduce the magnetic potential given by Eq.(\ref{konsimp}). Two
distinct cases may be considered, since a break in the
T$_{\textrm{d}}$ symmetry occurs: magnetization ``in-plane''
(occurring parallel to the interfaces)
${\bf{M}}_{\parallel}=M_x\hat{i}+M_y\hat{j}$, and ``perpendicular
to the plane'', ${\bf{M}}_{\perp}=M_z\hat{k}$. The latter is
assumed to be in the growth direction, here considered as the
z-axis. Notice that in the presence of a confining potential
created by the interfaces, $V_c(z)$, the operators $\hat J_x$ and
$\hat J_y$ no more commute with the LK Hamiltonian. In bulk,
however, within the homogeneous magnetization approach, there is
no distinction between these two cases. Once we are not trying to
explain the origin of the ferromagnetic order in these systems,
the occurrence of the magnetization being ``in plane'' or
``perpendicular-to-the-plane'' is assumed to be provided by an
external weak magnetic field, which does not interfere on the
electronic structure, directly.

For the sake of obtaining the LK matrix for heterostructures, it
is necessary to calculate first the matrix elements $<j,m\vert
V_{mag}\vert j',m'>$ for each constituent DMS layer, in bulk. This
is easily done by observing that
\begin{eqnarray}
{\bf{M}}_{\perp}\cdot\vec\sigma\vert \uparrow >&=&M_z\vert\uparrow >,\label{Mzsig}\\
{\bf{M}}_{\perp}\cdot\vec\sigma\vert \downarrow
>&=&-M_z\vert\downarrow
>,
\end{eqnarray}
and
\begin{eqnarray}
{\bf{M}}_{\parallel}\cdot\vec\sigma\vert\uparrow >&=&(M_x+iM_y)\vert
\downarrow >\equiv M_{+}\vert \downarrow >,\\
{\bf{M}}_{\parallel}\cdot\vec\sigma\vert\downarrow
>&=&(M_x-iM_y)\vert \uparrow >\equiv M_{-}\vert \uparrow >, \label{Mxysig}
\end{eqnarray}
Making use of the approximation given in Eq.(\ref{konsimp}) and
the results in Eqs. (\ref{Mzsig}-\ref{Mxysig}), we have for the LK
matrix of $V_{mag}$ at the $\Gamma$ point:
\begin{equation}
\tilde{V}_{mag}=-\frac{x}{6}N_0\beta \left( {\begin{array}{cccccc}
3M_z & 0 & i\sqrt{3}M_- & 0 & \sqrt{6}M_- & 0 \\
0 & -3M_z & 0 & -i\sqrt{3}M_+  & 0 & -\sqrt{6}M_+  \\
-i\sqrt{3}M_+  & 0 & M_z & 2iM_- & 2\sqrt{2}iM_z & -\sqrt{2}M_- \\
0 & i\sqrt{3}M_-  & -2iM_+ & -M_z &\sqrt{2}M_+  &  -2\sqrt{2}iM_z \\
\sqrt{6}M_+ & 0 &  -2\sqrt{2}iM_z & \sqrt{2}M_-  & -M_z & iM_-  \\
0 & -\sqrt{6}M_-  &  -\sqrt{2}M_+  & 2\sqrt{2}iM_z& -iM_+ &M_z
\end{array}}
\right)
\end{equation}

During the last years the LK model has been adapted to quantum
wells and superlattices (SL), as described in Refs.
\onlinecite{Ender,gui,sara}. We adopt that approach, using a
super-cell model. This means that we consider an unit cell
consisting of the active region plus a thick insulator layer. The
number of DMS layers in the unit cell can be varied at will.  We
assume, then, an infinite SL in the [001] direction. The multiband
effective-mass equation (EME) is represented with respect to plane
waves with wavevectors $K=(2\pi/a)l$ ($l$ an integer and $a$ the
SL period) equal to the reciprocal SL vectors. A detailed
description of the method can be found in Ref.
\onlinecite{sarath}. The rows and columns of the $6\times 6$ LK
Hamiltonian relate to the Bloch-type eigenfunctions $\vert
j,m_j,\vec{k}>$ of the $\Gamma_8$ heavy-hole bands, and the
$\Gamma_7$ spin-orbit-hole band; $\vec{k}$ denotes a vector of the
first Brillouin zone. Expanding the EME with respect to plane
waves $<z\mid K>$ means representing this equation in terms of the
Bloch functions $<{\bf x}\mid j,m_j,\vec{k}+K{\bf e}_z>$. For a
Bloch function $<z\mid E,\vec{k}>$ of the SL corresponding to
energy $E$ and wavevector $\vec k$, the EME takes the form:
\begin{eqnarray}
\nonumber \sum_{j',m_j',K'}<j,m_j,\vec{k},K\mid
T+H_S+V_{het}+V_C+V_{xc}+V_{mag}\mid
j',m_j',\vec{k},K'>\times\\
<j',m_j',\vec{k},K'\mid E,\vec{k}>=
E(\vec{k})<j,m_j,\vec{k},K\mid E, \vec{k}>, \label{secular}
\end{eqnarray}
where $T$ is the unperturbed kinetic energy term generalized for a
heterostructure, \cite{saraapl} $H_S$ is the strain energy term
originating from the lattice mismatch, $V_{het}$ is the square
potential due to the difference between energy gaps, $V_{xc}$ is
the exchange-correlation potential, and $V_C$ is the sum of the
Hartree potential with the ionized acceptor potential. Finally,
$V_{mag}$ is given by Eq.(\ref{konsimp}), for each material. The
Luttinger parameters and the other terms appearing in the secular
equation are to be taken for each epitaxial layer of the SL.
\cite{lutpar} For instance, in the case of the magnetic
interaction we have:
\begin{equation}
<j,m,\vec{k},K\vert V_{mag}\vert j',m',\vec{k}',K'>=
<\vec{k},K\vert \tilde{V}_{mag}^{jm;j'm'}g(K'-K)\vert
\vec{k}',K'>, \label{defkkp}
\end{equation}
where the integral
\begin{equation}
g(K'-K)=\frac{1}{d}\int_0^d e^{-iKz} g(z)e^{iK'z}
\end{equation}
is performed in a DMS layer of width $d$.

The self-consistent potentials and the charge densities are
obtained by solving the multiband EME equation and the Poisson
equation:
\begin{equation}
<j,m_j,\vec{k},K\mid V_C\mid j',m_j',\vec{k},K'>=\frac{4\pi
e^2}{\kappa}\frac{1}{\vert K-K'\vert^2}<K\mid \rho^++\rho^-\mid
K'>\delta_{j,j'}\delta_{m_j,m'_j},
\end{equation}
where $\kappa$ is the dielectric constant of the host, and
$\rho^+$ and $\rho^-$ are the density of charge of holes and
acceptors, respectively, expressed in plane-wave representation.

\section{Results}

The DMS layers work effectively as barriers or wells for spins
parallel (up) and antiparallel (down) to the local average
magnetization, depending on the sign of $N_0\beta$ for valence
band, and $N_0\alpha$ for conduction band. These DMS layers are
assumed to be ferromagnetic and metallic, with a 3-D equivalent
hole density $p= 1$ x $10^{20}$ cm$^{-3}$, a substitutional Mn
concentration of $5\%$, at temperature T= 0 K, and an average
magnetization $\langle M \rangle$=5/2.

In Fig. \ref{fig2} we present (a) the valence band structure (hole
binding energy)  and (b) potential profiles for a  system
consisting of six DMS layers, with $d_1$=20 {\AA} and $d_2$=30
{\AA}. Energies are reckoned from the top of the Coulomb barrier,
as in Ref. \onlinecite{sarath}. The magnetization is assumed to be
in the z-direction. The subbands are hybrid states, since they are
a mixing of all kinds of holes. However, at the $\Gamma$-point the
lowest lying states have a dominant component. For instance, the
first three states are almost entirely  heavy holes "down". The
mixing becomes stronger in the more excited states, and as we go
out of the $\Gamma$-point. In Fig. \ref{fig2} we named the band by
its dominant component  at the $\Gamma$-point. Here, ``up'' and
``down'' refer to the sign of $m_j$, the z-component of the total
angular momentum. In other words, ``up'' means parallel to the
average magnetization, while ``down'' means anti-parallel. The
$\Gamma$-$\Delta$ ($\Gamma$-Z)- line corresponds to wave vectors
$\vec{k}$ perpendicular (parallel) to the SL axis. The Fermi
energy is also indicated. Strong non-parabolicity arises in the
subbands along the ($\Gamma$-$\Delta$)-line, which leads to
remarkable anti-crossing behavior. We also observe that several
levels are occupied, most of them are heavy and light hole down.
It is possible to understand this behavior by observing Fig.
\ref{fig2}(b), which shows the self-consistent hole band potential
profile for each carrier. The confinement for heavy and light
holes down are deeper than the cases where the z-components of the
total angular momentum are up.

Fig. \ref{fig3} shows the carriers distribution for the structure
with six DMS layers described in Fig. \ref{fig2}. The results are
shown for lateral GaAs layers of width  $s=$0, 10, 20, 30, 40, 50
and 60~\AA. We notice that, except for the heavy hole with
$m_j=3/2$ (up), carriers concentrate on the DMS layers.   The
heavy holes up concentrate in the non-magnetic region as a
consequence of the strong magnetic repulsion. After $s=50$\AA\ the
carriers distribution becomes independent of the lateral GaAs
widths.

Fig. \ref{fig4} shows the carriers density for active regions
consisting of 1 to 9 DMS layers. We fixed $d_1$=20 {\AA} and
$d_2$=30 {\AA}. The charge density is plotted for each component.
As before, ``up'' and ``down'' refer to the sign of the $m_j$
component of the total angular momentum. These results are
consistent with the electronic structure shown in Fig. \ref{fig2},
since the lowest levels are mostly heavy and light holes down.
Observe that in all multilayered structures (2 to 9 DMS layers)
the charge is concentrated almost entirely inside the DMS layers.
Again, the exception is the heavy hole up component, whose density
is higher in the non-magnetic regions for the reasons explained
above.

Up to now  the distribution of charge in the multilayered
structure has been shown in terms of carriers total angular
momentum components. For the sake of completeness, it is also
interesting to know how the charge is distributed in terms of the
spin polarization. This can be easily obtained, since the light
holes are mixed states of up and down spins with defined
probabilities. The results for these multilayers are shown in
Fig.\ref{fig5} in terms of the spin components of the charge
density. There is a strong spin polarization, dominated by heavy
and light holes down spins.

In Fig. \ref{fig6} we analyze the dependence of the effective
two-dimensional hole concentration, N$_{2D}$, which is the
integrated 3-D density on the z-direction for each component. The
DMS layers widths are changed, while keeping fixed the width of
the GaAs layers, $d_2$=30 {\AA}. The calculation is performed for
6 DMS layers. For the heavy and light hole down, the 2-D densities
are much higher then those associated to positive values of $m_j$.
The difference is one order of magnitude, in the case of heavy
holes, due to its higher effective mass and profile potential that
provide a higher occupation inside the DMS layers.

Finally, in Fig. \ref{fig7} we show N$_{2D}$ as a function of the
non-magnetic layer width, $d_2$, with $d_1$=20 {\AA}, for each
carrier, heavy and light hole (down and up). The calculation is
performed for 4, 5, 6 and 7 DMS layers.

\section{Conclusions}

In summary, we have investigated the electronic structure of
Ga$_{0.95}$Mn$_{0.05}$As/GaAs multilayers by using a super-cell
model in the framework of the Luttinger-Kohn ${\bf k. p}$
approximation. The unit in the super-cell is  a
AlAs/n(Ga$_{0.95}$Mn$_{0.05}$As/GaAs)/AlAs structure, with $n$
being the number of DMS layers, grown in the [100] direction. The
DMS layers are assumed to be metallic and ferromagnetic, at T=0K.
A small magnetic field is applied in the growth direction to
guarantee the magnetization to be perpendicular to the plane, but
it does not affect the electronic structure. Several subbands are
occupied. They are mostly heavy holes with $m_j=-3/2$, the carrier
density being higher in the DMS layers. However, a non-negligible
part of carriers are light holes with $m_j=-1/2$, also
concentrated on the same region as the heavy holes. Actually, what
is observed is the appearance of two different channels, one with
the z-component of the total angular  momentum aligned with the
average magnetization, which is highly concentrated on the DMS
layers, and another, with opposite orientation whose distribution
is more concentrated on the non-magnetic layers of the active
region, the latter channel being much less dense. These results
are in qualitative agreement with those obtained in Ref.
\onlinecite{sanvito} showing the electronic structure calculations
in digital ferromagnetic heterostructures of  GaAs/(Ga,Mn)As and
AlAs/(Ga,Mn)As/GaAs using density-functional theory in the local
spin-density approximation. As a consequence of both magnetic and
Coulomb interactions - here included the hole-hole interaction -
the spin-polarized charge tends to be slightly non-periodically
distributed as the number of DMS layers increases. Carriers
polarized anti-parallel to the magnetization tend to concentrate a
little more at the borders of the structure, while carriers
polarized in the opposite direction tend to concentrate on the
middle.

From the point of view of the observation of a ferromagnetic
multilayer with a high transition temperature (we recall that we
have assumed in our calculation T=0K), it is interesting to have a
strong ``spin-polarized'' charge density in the DMS layers, while
keeping some charge in between to guarantee the inter-layer
interaction.\cite{gr1,gr2} However, for obtaining a high mobility
spin-polarized current, a structure should be grown in which the
in-plane transport is realized with spin-polarized carriers
concentrated in a region of high mobility, away from the
scatterers. Therefore, the ideal distribution of charge and spin
depends on the purpose of the structure. The results we have
obtained points to the possibility of engineering the
spin-polarized charge distribution by the right choice of the
magnetic layers and the band mismatches with the non-magnetic
spacers.

Acknowledgments: Work supported by CNPq (NanoSemiMat network), and
FAPESP. ICCL thanks the LNMS group for its generous hospitality.

\newpage
\begin{figure}
\caption{The model structure: the DMS layers of width $d1$ are
separated by non-magnetic GaAs layers of width $d_2$. At the left
of the first DMS layer and at the right of the last one, GaAs
lateral layers of width $s$ complete the ``active'' parts. The
``active'' parts are separated, themselves, by thick AlAs
barriers.} \label{fig1}
\end{figure}

\begin{figure}
\caption{(a)Valence band structures for GaAs/GaAs:Mn  multiple QWs
along high symmetry lines $\Gamma$-$Z$ and $\Gamma$-$\Delta$, with
$d_1= 20$ \AA~and $d_2= 30$ \AA. We named the band by its dominant
component  at the $\Gamma$-point. Solid lines correspond to
subbands which, at the $\Gamma$-point, are mostly heavy holes
down, dashed lines to light hole down, dotted lines to heavy hole
up, and dotted-dashed lines to light holes up. The Fermi energy is
also indicated (short-dashed line). (b) the corresponding
potential profile for each carrier, down and up. Energies are
reckoned from the top of the Coulomb barrier. } \label{fig2}
\end{figure}

\begin{figure}
\caption{Density distribution of heavy and light holes for six DMS
layers with $d_1=20$\AA, $d_2=30$\AA, and lateral GaAs widths
$s=0, 10, 20, 30, 40, 50$ and 60\AA.} \label{fig3}
\end{figure}

\begin{figure}
\caption{Particle density of heavy and light holes for a structure
consisting from 1 to 9 DMS layers. Widths $d_1$ and $d_2$ as in
Fig.{\protect{\ref{fig1}}}.} \label{fig4}
\end{figure}

\begin{figure}
\caption{Same as in Fig. {\protect{\ref{fig3}}}, for each carrier
spin component.} \label{fig5}
\end{figure}

\begin{figure}
\caption{N$_{2D}$  as a function of the DMS layer width $d_1$. The
width of the non-magnetic layers are fixed, $d_2= 30$ \AA.}
\label{fig6}
\end{figure}

\begin{figure}
\caption{N$_{2D}$  as a function of the non-magnetic layer width
$d_2$, for 4, 5, 6 and 7 DMS layers. The widths of the DMS layers
are fixed, $d_1= 20$ \AA.} \label{fig7}
\end{figure}

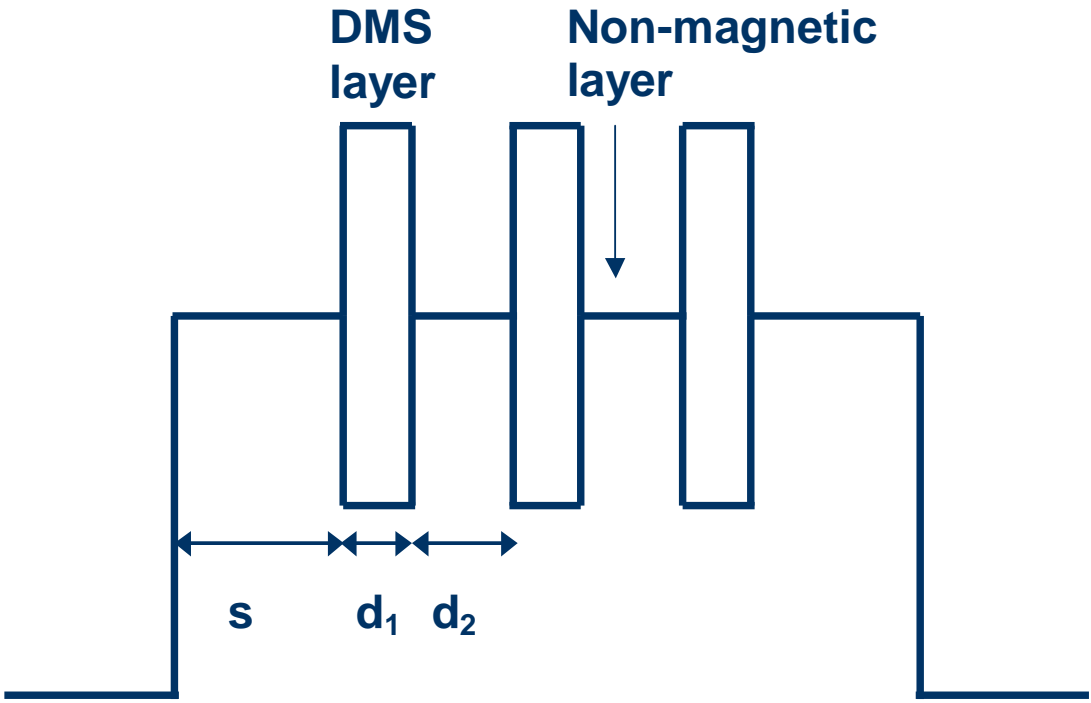
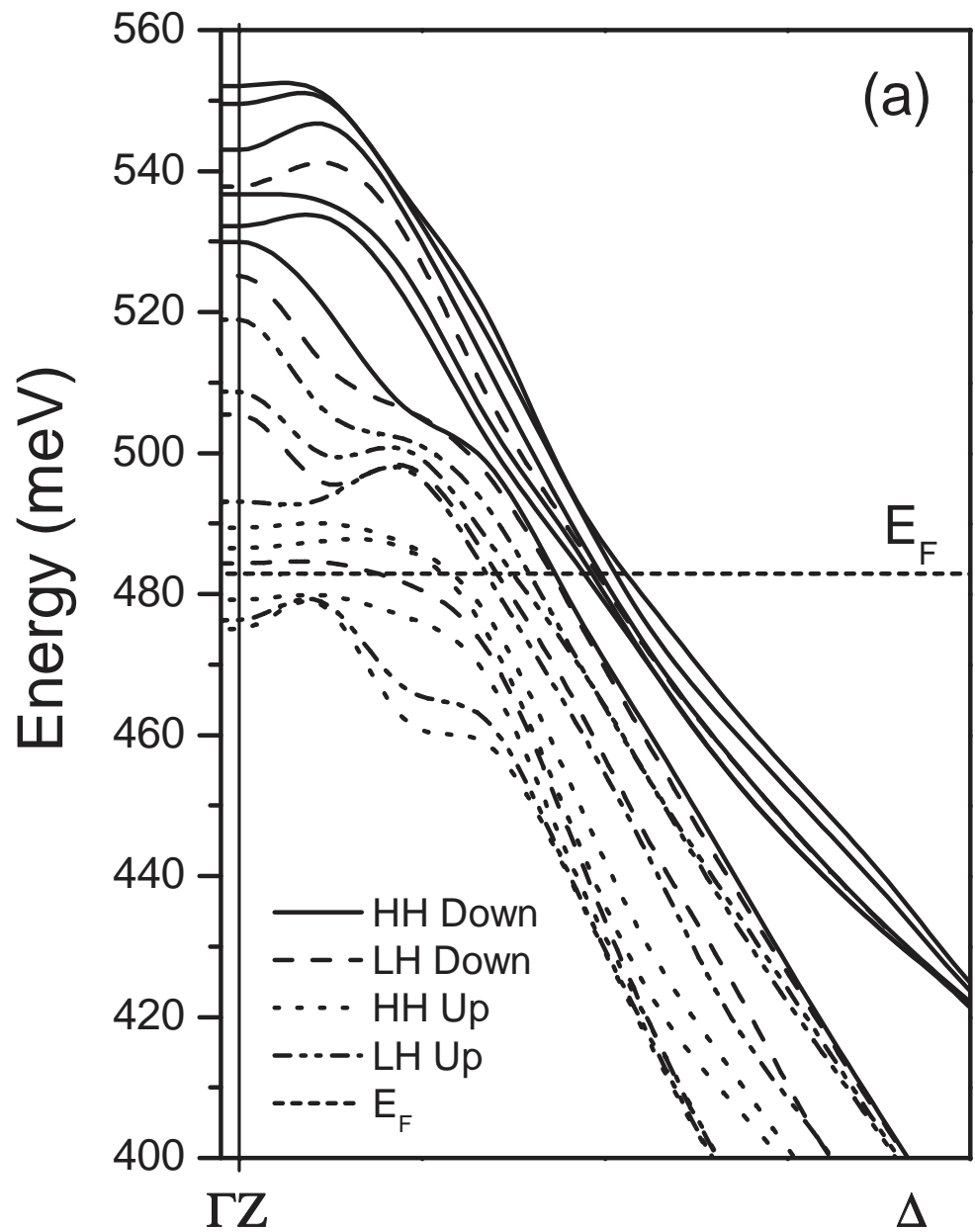
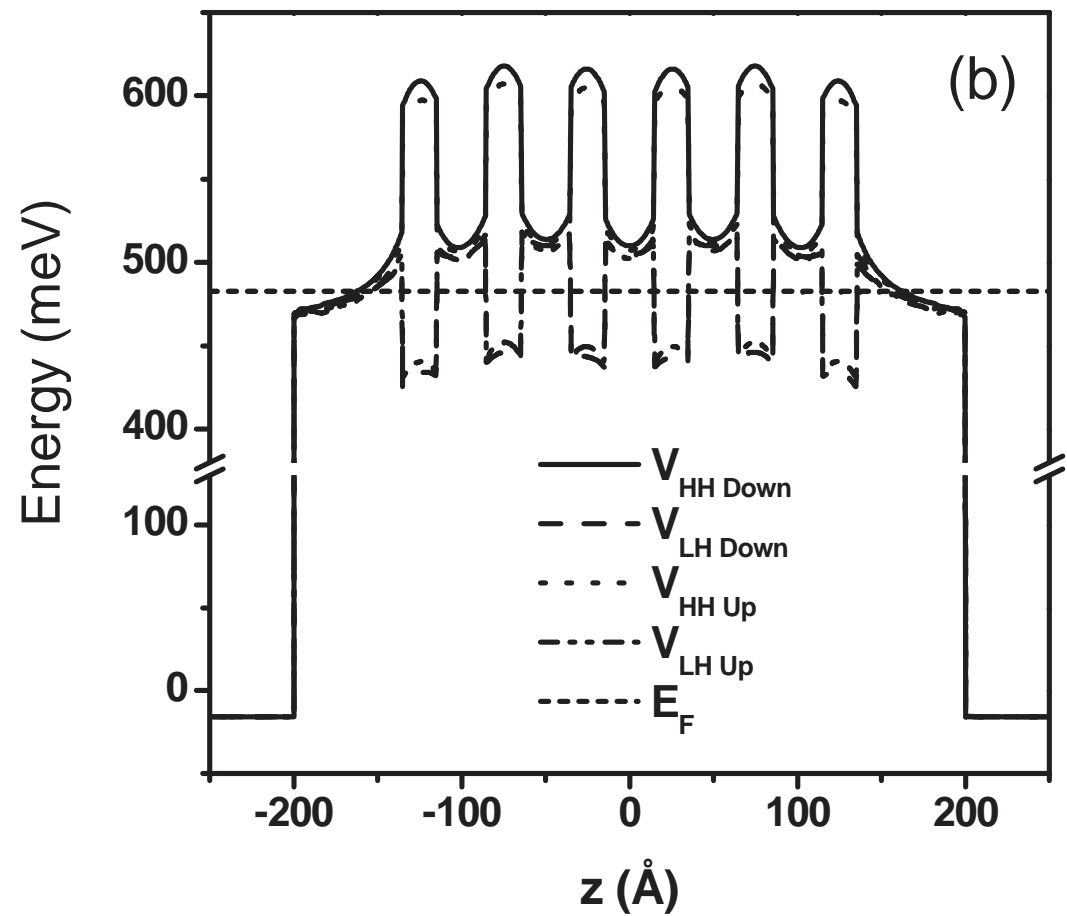
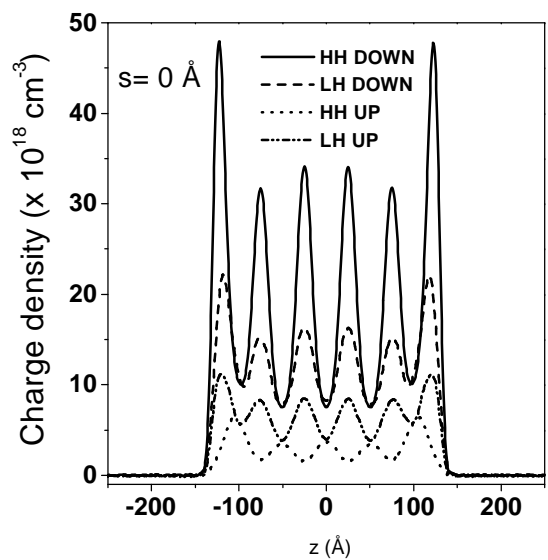
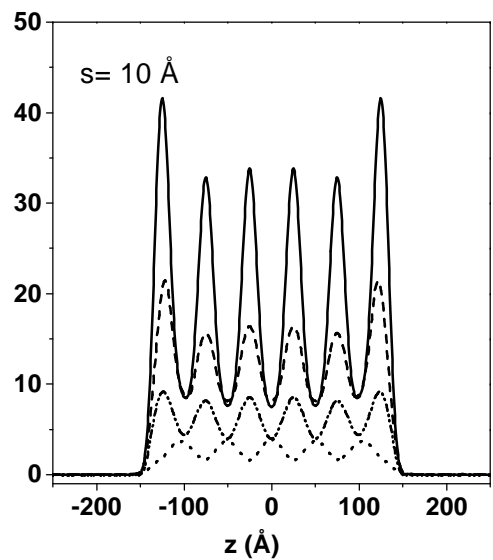
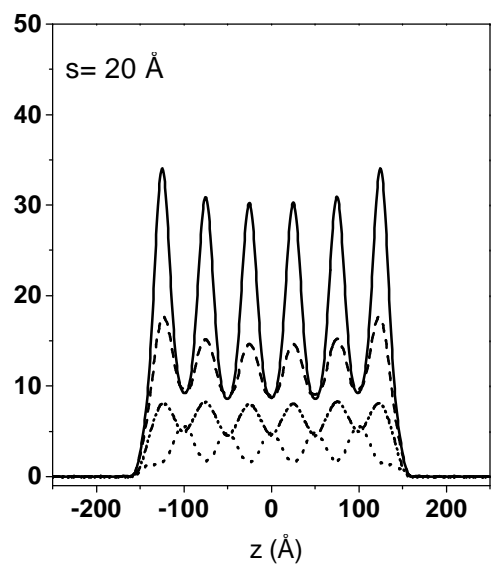
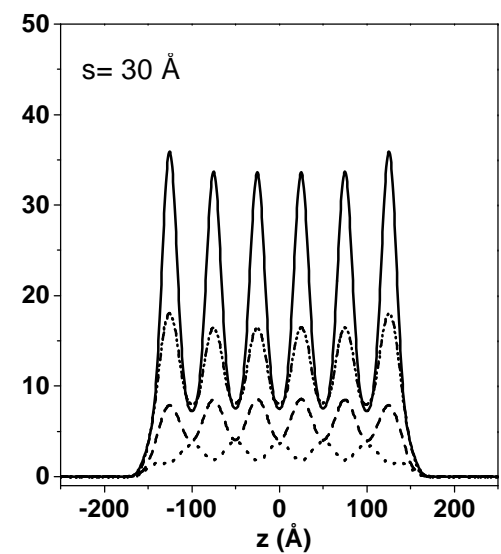
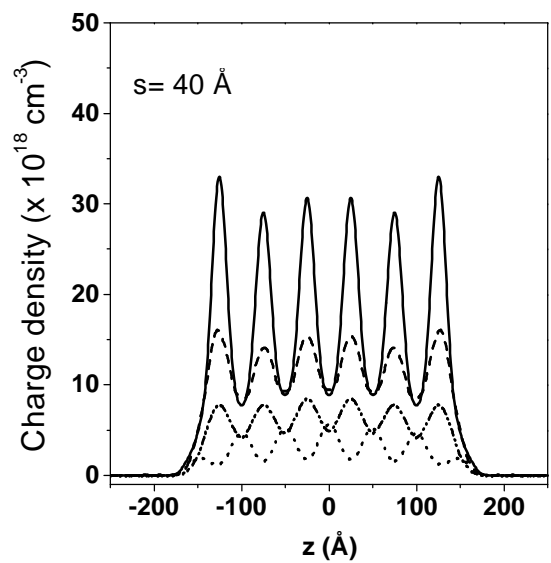
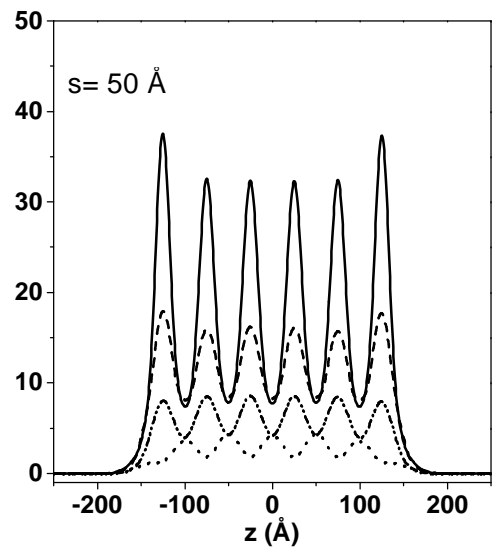
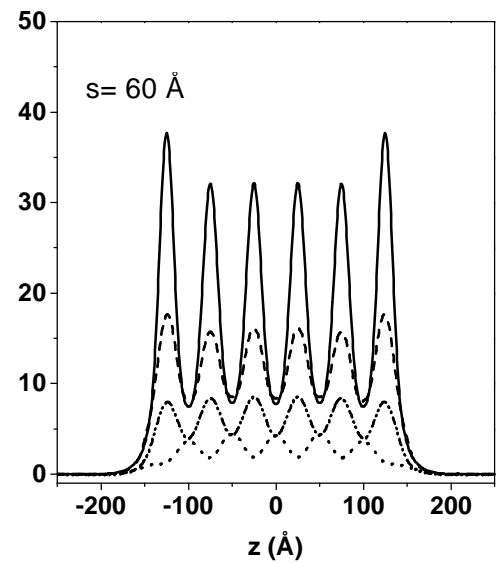
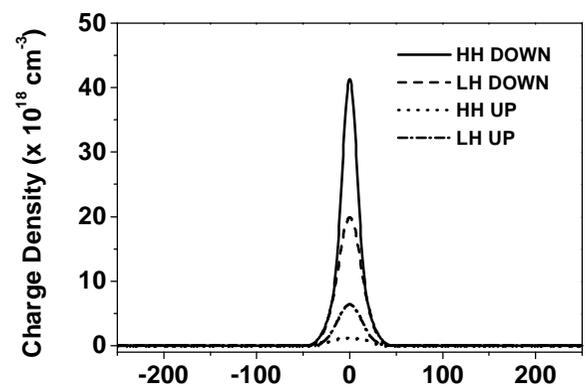
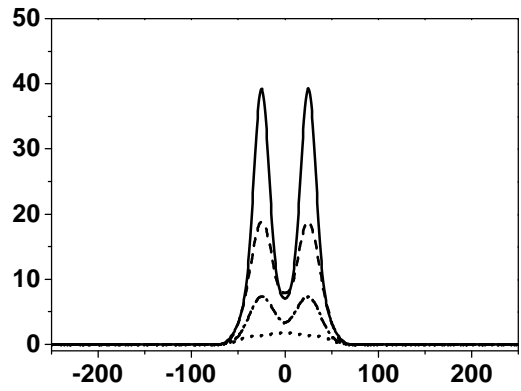
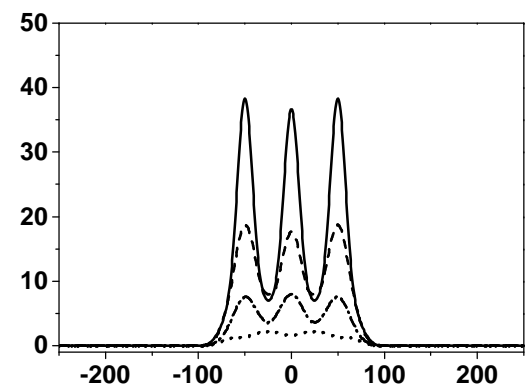
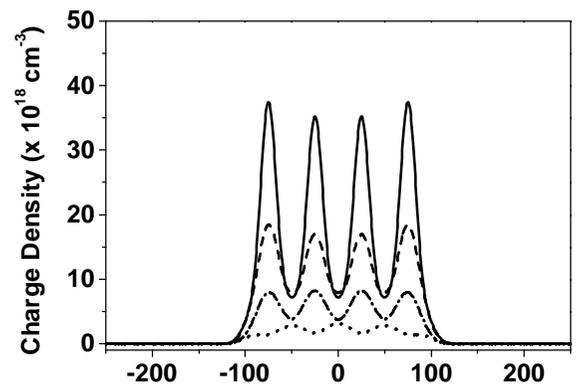
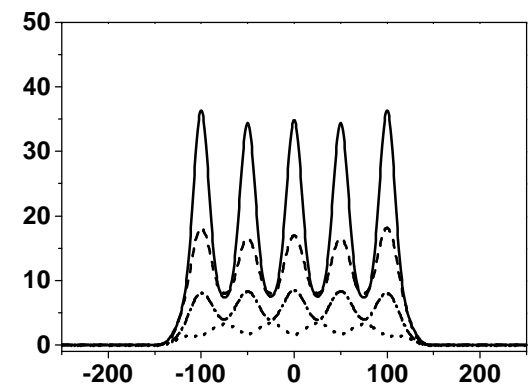
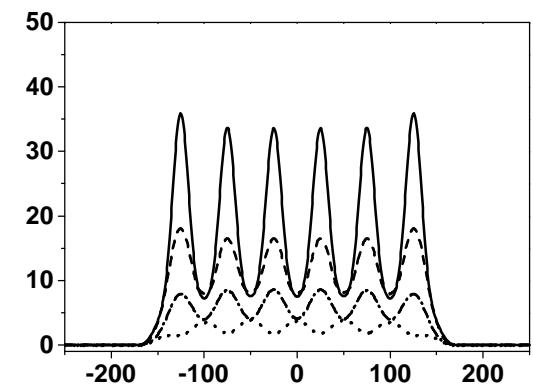
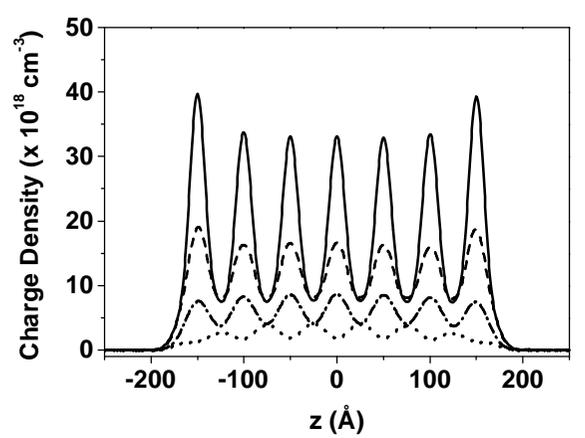
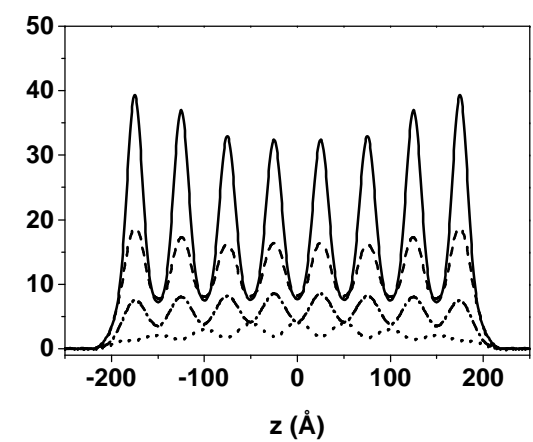
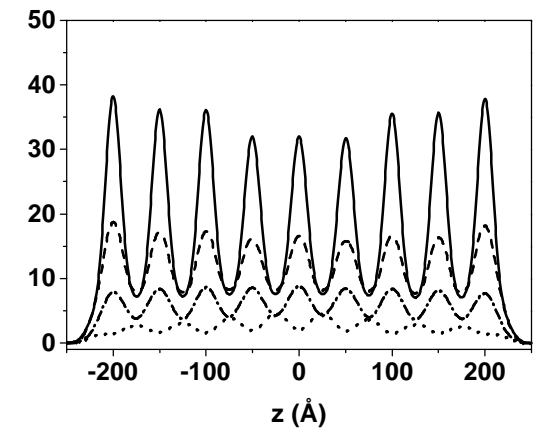
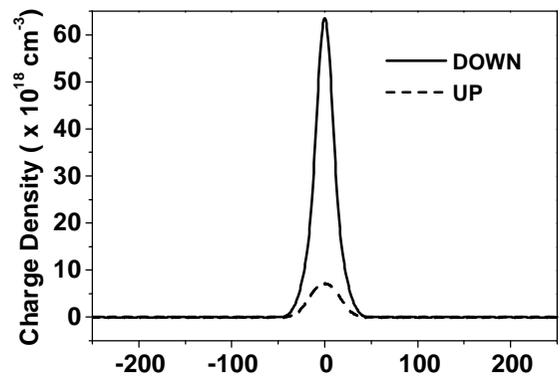
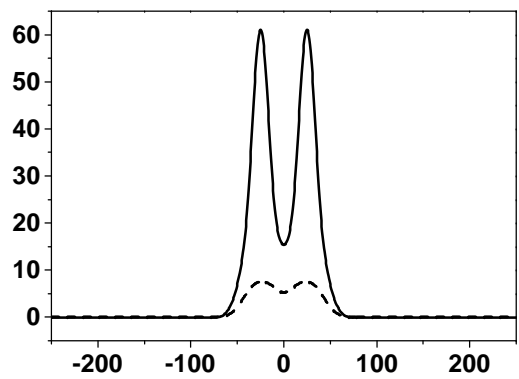
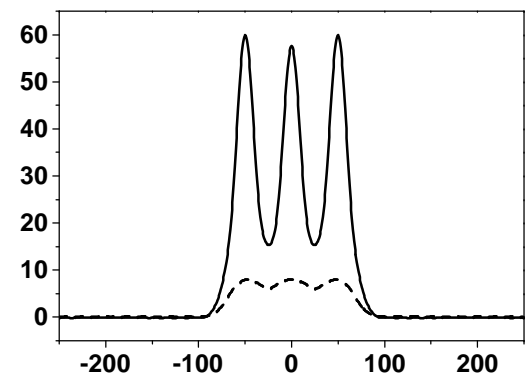
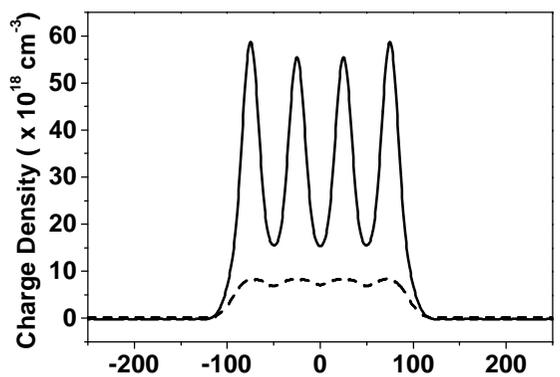
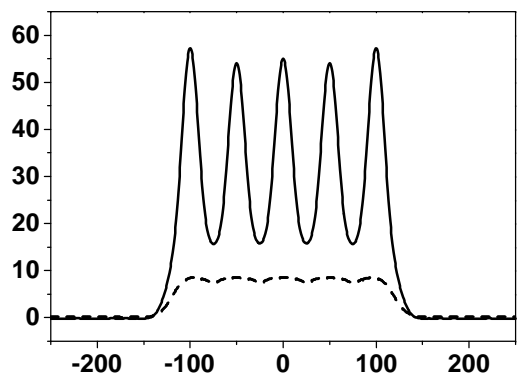
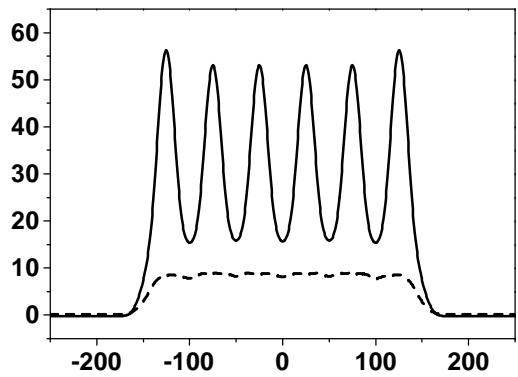
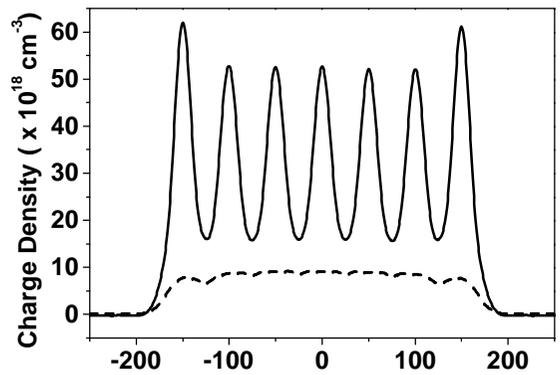
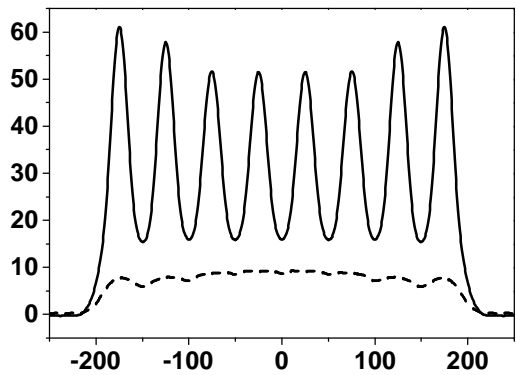
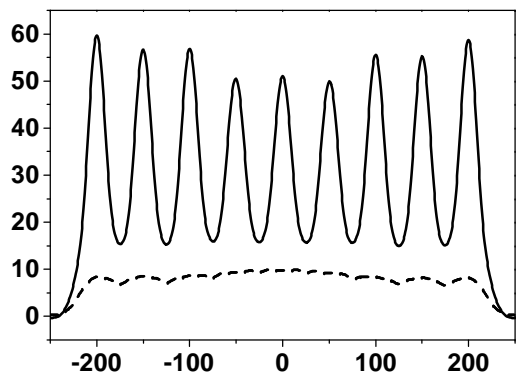
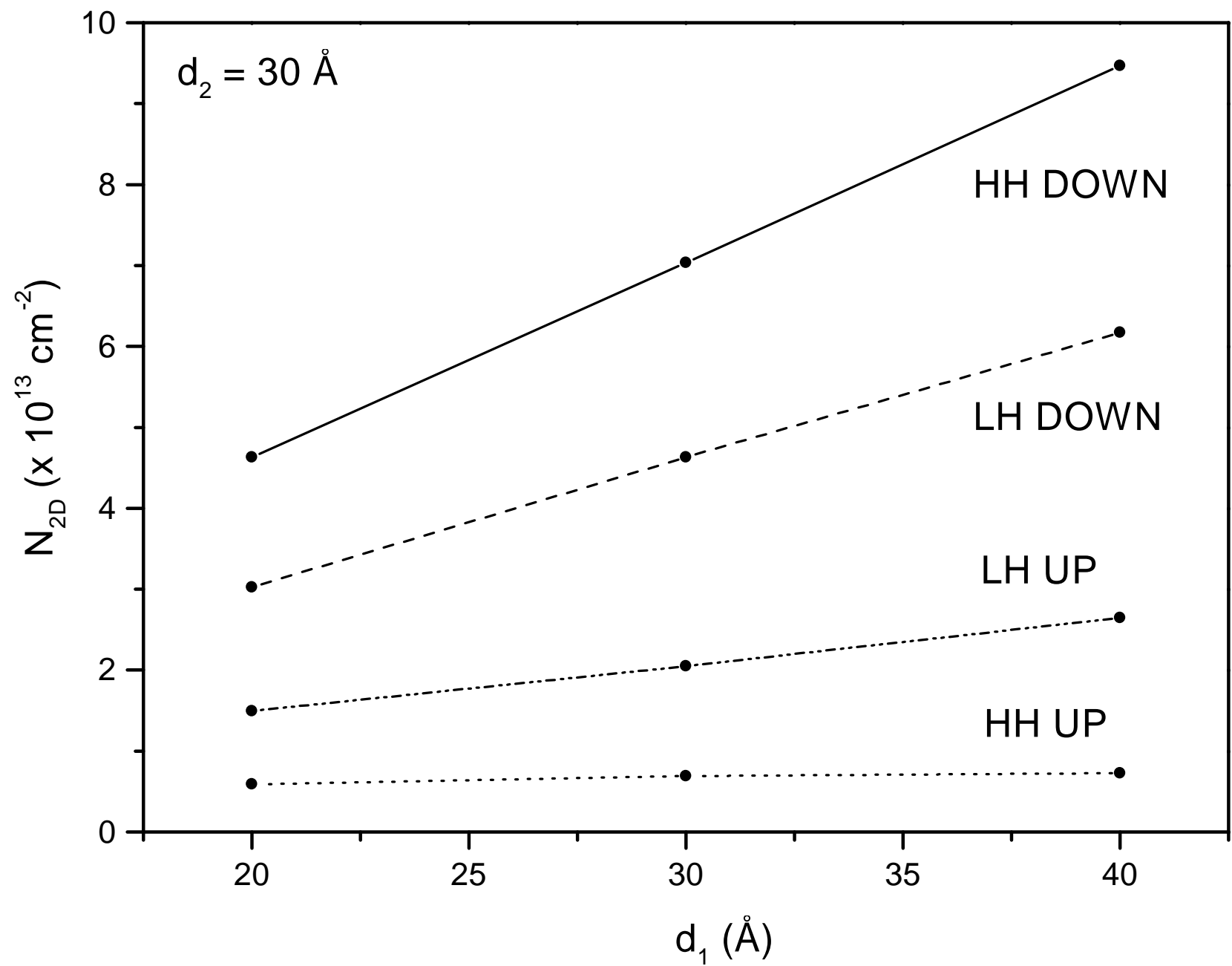
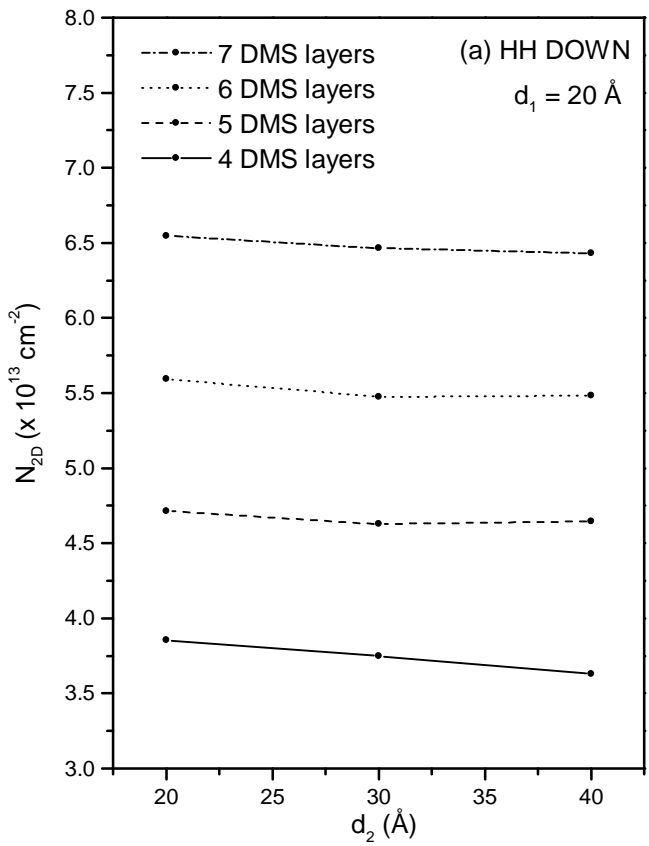
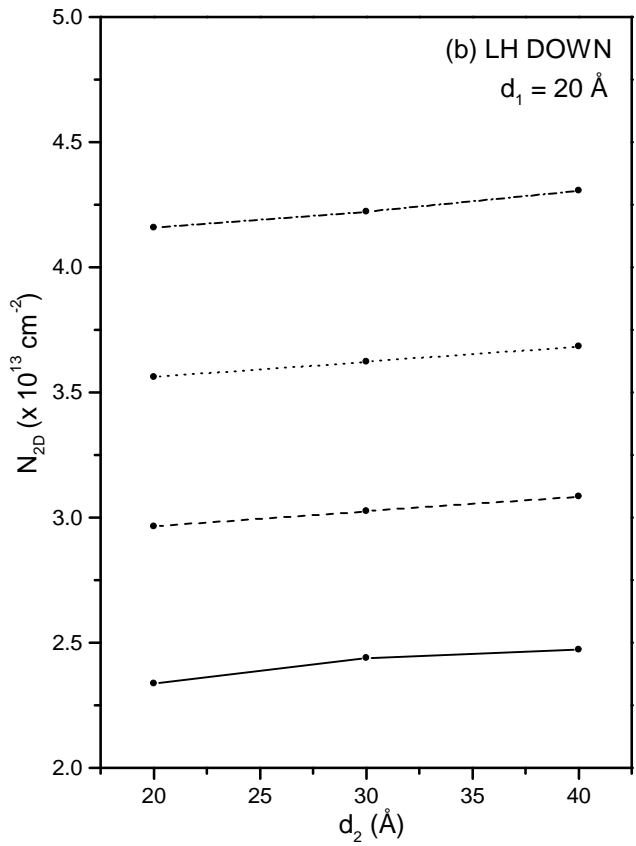
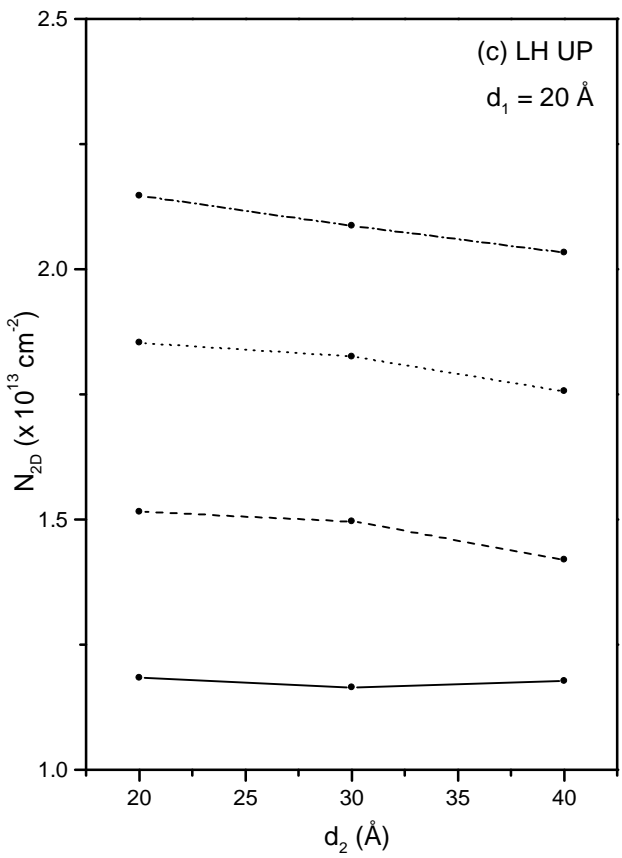
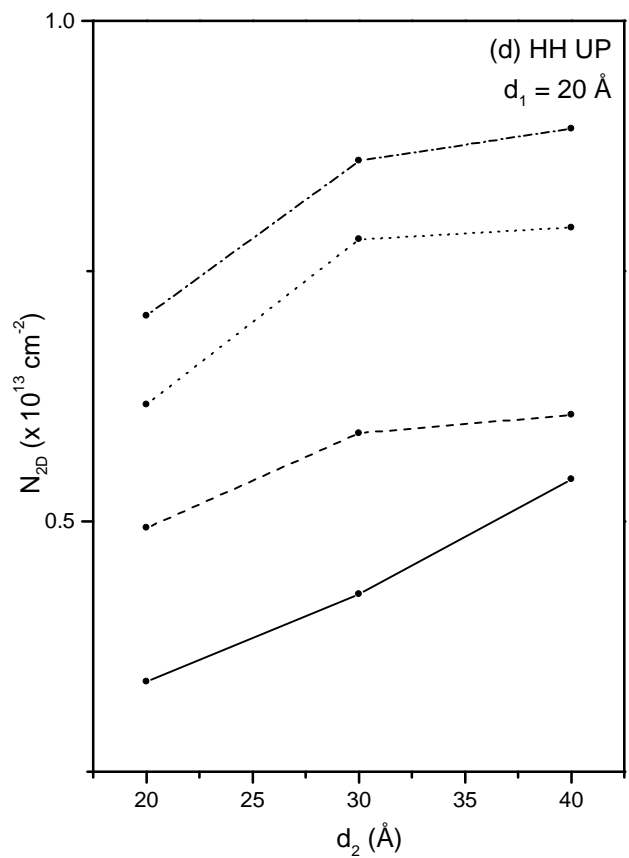
\end{document}